\documentclass[12pt,preprint]{aastex}
\shorttitle{RGBbump47Tuc}
\shortauthors{Nataf et al.}
\usepackage{natbib}
\usepackage{float}
\usepackage{array,longtable}

\title{The Gradients in the 47 Tuc Red Giant Branch Bump and Horizontal Branch are Consistent With a Centrally-Concentrated, Helium-Enriched Second Stellar Generation}
\author{D. M. Nataf\altaffilmark{1}, A.  Gould\altaffilmark{1}, M.H. Pinsonneault\altaffilmark{1}, P.B. Stetson\altaffilmark{2}}
\altaffiltext{1}{Department of Astronomy, Ohio State University, 140 W. 18th Ave., Columbus, OH 43210}
\altaffiltext{2}{Herzberg Institute of Astrophysics, National Research Council
Canada, 5071 West Saanich Road, Victoria, BC V9E 2E7}
\email{nataf@astronomy.ohio-state.edu}

\begin{document}

\begin{abstract}
We combine ground and space-based photometry of the Galactic globular cluster 47 Tuc to measure four independent lines of evidence for a helium gradient in the cluster, whereby stars in the cluster outskirts would have a lower initial helium abundance than stars in and near the cluster core. First and second, we show that the red giant branch bump (RGBB) stars exhibit gradients in their number counts and brightness. With increased separation from the cluster center, they become more numerous relative to the other red giant (RG) stars. They also become fainter. For our third and fourth lines of evidence, we show that the horizontal branch (HB) of the cluster becomes both fainter and redder for sightlines farther from the cluster center. These four results are respectively detected at the 2.3$\sigma$, 3.6$\sigma$, 7.7$\sigma$ and 4.1$\sigma$ levels. Each of these independent lines of evidence is found to be significant in the cluster-outskirts; closer in, the data are more compatible with uniform mixing. Our radial profile is qualitatively consistent with but quantitatively tighter than previous results based on CN absorption. These observations are qualitatively consistent with a scenario wherein a second generation of stars with modestly enhanced helium and CNO abundance formed deep within the gravitational potential of a cluster of previous generation stars having more canonical abundances.
\end{abstract}

\section{Introduction}
\label{sec:Introduction}
47 Tuc (NGC 104) is among the most massive globular clusters in the Galaxy and is thus one of the most powerful laboratories to investigate the finer details of globular cluster formation and evolution. As the large stellar population renders any potential statistic more accessible, it is interesting that 47 Tuc is not among those globular clusters with more clearly delineated evidence for multiple populations \citep{2009AJ....138.1455B}. 

However, it has been known for several decades that the stars in the inner part of the globular cluster have stronger CN absorption \citep{1979ApJ...230L.179N,1990BAAS...22.1289P}. Recently, \citet{2010MNRAS.408..999D} have argued that this chemical gradient is due to the presence of multiple generations of stars, with later generations being helium and CN enhanced by the ejecta of first-generation asymptotic giant branch (AGB) stars. They found strong evidence of a helium-spread in the morphology of the subgiant branch (SGB) and horizontal branch (HB) stars. Their work followed an investigation by \citet{2009ApJ...697L..58A}, who used {\it Hubble Space Telescope (HST)} data to measured the color widths of the cluster main sequence, which they argued could be explained by a spread of ${\Delta}Y \sim 0.027$. If the helium-enhancement is due to a second generation, and if the second generation is indeed more centrally concentrated, as suggested by the CN band strengths and dynamical arguments \citep{2008MNRAS.391..825D}, one should expect a higher helium abundance in the center. It has recently been posited that the presence of multiple generations differing in properties such as initial helium abundance and the relative abundances of sodium and oxygen are in fact a ubiquitous property of globular clusters \citep{2010A&A...516A..55C}.

In this paper we test the hypothesis of a helium gradient in 47 Tuc using four methods that are rooted in the properties of two densely-populated phases of post main-sequence stellar evolution, the red giant branch bump (RGBB) and the HB. The RGBB phase occurs during the first ascent of the red giant branch. As the hydrogen burning shell expands, it eventually comes into contact with the convective envelope \citep{1997MNRAS.285..593C}. This increase in fuel causes the star to become fainter as the fuel is used up before becoming brighter again, effectively crossing the same luminosity three times, leading to a ``bump'' in the luminosity function. This bump is most populated and thus more measurable in metal-rich clusters such as 47 Tuc \citep{1999ApJ...518L..49Z,2001ApJ...546L.109B,2003A&A...410..553R,2010ApJ...712..527D,2011A&A...527A..59C}. 

Stellar evolution predicts the RGBB lifetime to be significantly shortened for increased initial helium abundance \citep{2001ApJ...546L.109B,2010ApJ...712..527D,2010arXiv1011.4293N}. If this stellar theory prediction is correct, and the hypothesis of a centrally-concentrated, helium-enhanced population in 47 Tuc is correct as well, then RGBB stars should be less prominent relative to the remaining RG stars closer to the cluster center. We detect a variation in the equivalent width (EW) of the RGBB at the $\sim$2.3$\sigma$ level. For our second test, we also show that the RGBB stars are fainter with increasing distance from the cluster center, a detection made at the $\sim$3.6$\sigma$ level. 

We also investigate the cluster HB stars. We show that the HB stars are both fainter ($\sim$7.7$\sigma$) and redder ($\sim$4.1$\sigma$) farther from the cluster center, though the gradient levels off in the inner $\sim$200$\arcsec$. This is the expected trend for increased helium in stellar models, as seen for example in Figure 1 of \citet{2010MNRAS.408..999D}, which was produced using the code ATON 2.0 \citep{1998A&A...334..953V,2009A&A...499..835V}. This gradient is already known from \citet{1997AJ....114.1051B}, who showed with high confidence that HB stars with strong CN absorption were $\sim$0.04 mag brighter. They argued that this effect could be explained by either a small difference in the core mass of helium burning stars or a small difference in the initial helium abundance of those brighter, CN-enhanced HB stars.

We have conducted these experiments by combining two independent datasets, a space-based dataset toward the cluster center \citep{2007AJ....133.1658S} and a comprehensive ground-based dataset  for the remaining sightlines \citep{2000PASP..112..925S}. The latter contains $\sim$60\% of the stellar sub-populations studied in this paper. The two datasets respectively include $\sim$500 and $\sim$700 HB stars, as well as $\sim$120 and $\sim$150 RGBB stars. For all four tests, we find significant detections of the trends expected from a helium gradient in the cluster outskirts, with no discernible trend within the cluster center. This is most consistent with a picture of two stellar generations that are evenly mixed within the cluster center, but with the second generation having a characteristic radius beyond which its numbers fall more rapidly. Our estimated transition radius of a few arcminutes is smaller than that derived from previous investigations of  CN absorption among cluster giants. \citet{1997AJ....114.1051B} found that the ratio of CN-strong to CN-weak stars was approximately constant up to $\sim$10-15 arcminute separations from the cluster center, beyond which the ratio fell rapidly.

\begin{figure}[H]
\begin{center}
\includegraphics[totalheight=0.7\textheight]{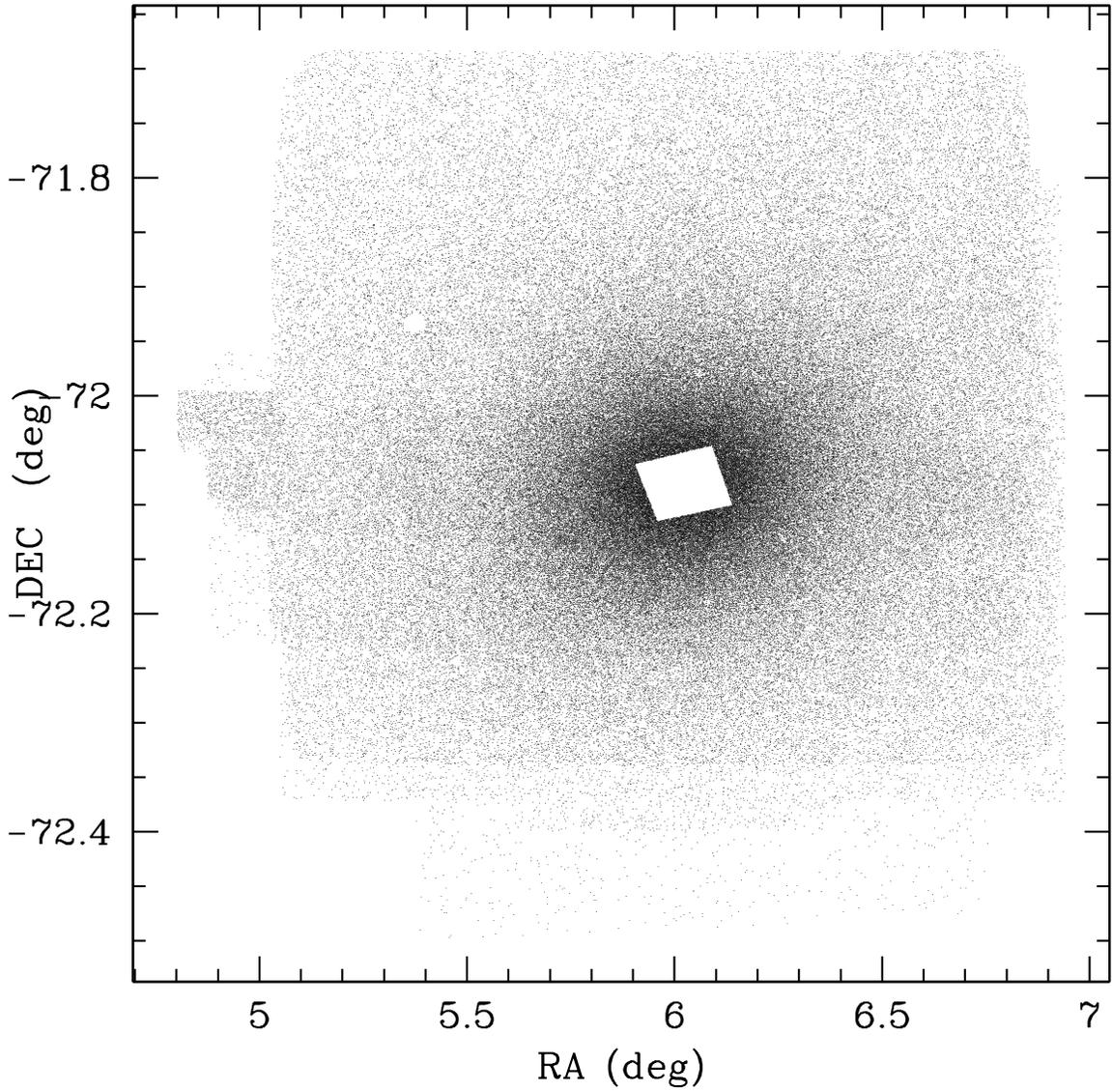}
\end{center}
\caption{The field of view (FOV) of observations used in this work. Point sources from the ground based photometry are shown as points. The inner parallelogram centered at (RA, DEC)$\sim$(6.023, -72.081) or $(\alpha, \delta) = (00:24:06, -72:04:52)$ corresponds to the FOV of the \textit{HST} dataset. }
\label{Fig:FieldOfView}
\end{figure}

\begin{figure}[H]
\includegraphics[totalheight=0.7\textheight]{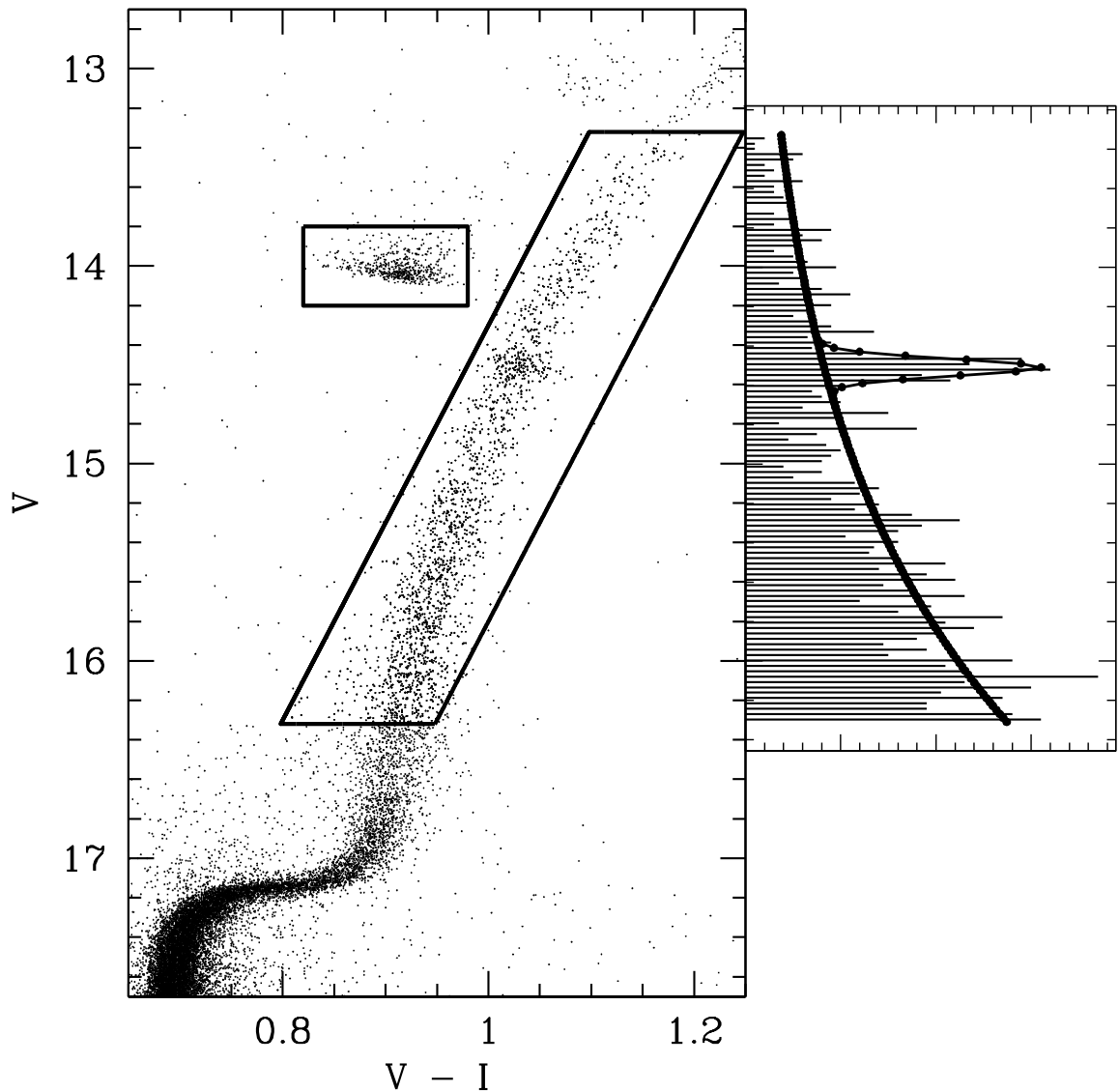}
\caption{LEFT: The color-magnitude diagram of the \textit{HST} dataset. The RG branch (including the RGBB), the HB, and the SGB are all contained within their respective color-magnitude selection boxes. RIGHT: Magnitude distribution of RG stars. The RGBB stands out as a prominent and significant peak at $V$ = 14.51.}
\label{Fig:CMDHubble}
\end{figure}


\begin{figure}[H]
\includegraphics[totalheight=0.7\textheight]{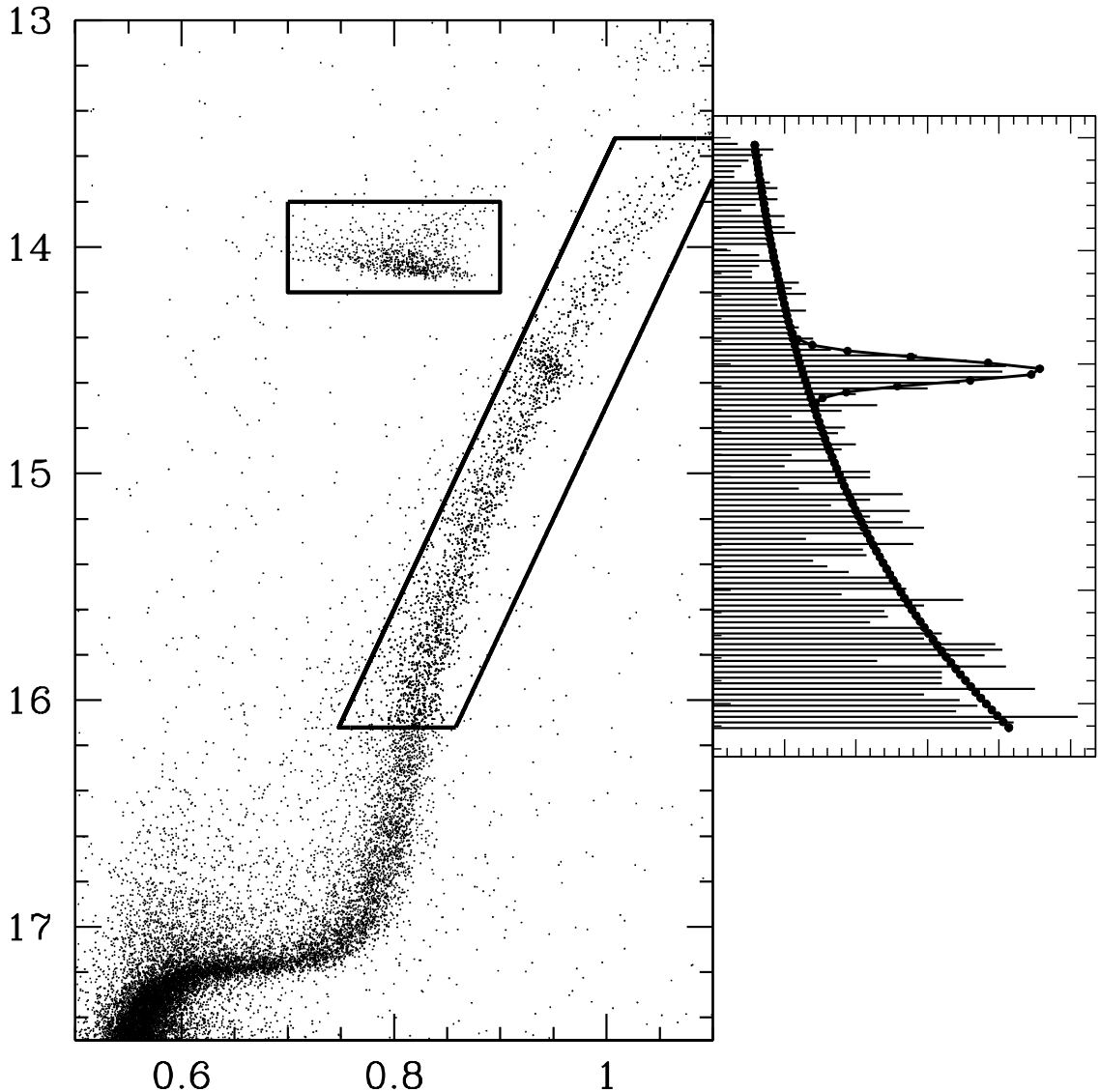}
\caption{Same as in Figure \ref{Fig:CMDHubble} but for the ground based data rather than the ACS data. The RGBB peak is detected at $V$ = 14.54. The difference of 0.03 mag with that found in the space-based dataset is very possibly due to differences in convention.}
\label{Fig:CMDGroundBased}
\end{figure}

\section{Data}
We make use of two different data sets in this study to maximize the available information and constrain the effect of any possible systematics. 

For the cluster center, we use photometry obtained with \textit{HST}'s Advanced Camera for Surveys (ACS) \citep{2007AJ....133.1658S}. The data were taken as part of an \textit{HST} treasury program to obtain high signal-to-noise ratio photometry  down the to the lower main sequence for a large number of Galactic globular clusters. We use $V_{ground}$ and $I_{ground}$ (hereafter ``$V$'' and ``$I$'') photometry, which were transformed from the original $F606W$ and $F814W$ photometry. Artificial star tests demonstrate that the photometry is expected to be very precise and complete at the brightness of the RGBB \citep{2008AJ....135.2055A}. 

We also use $U$, $B$, $V$, and $I$ observations that come from a database of original and archival observations \citep{2000PASP..112..925S}, which are calibrated on the \citet{1992AJ....104..340L} photometric system. These observations and the general properties of the 47 Tuc color-magnitude diagram (CMD) are described in \citet{2009AJ....138.1455B}. We make use of stars in these observations that are outside of the coordinate range observed by the ACS dataset. The 135 point sources that are located within 30$\arcsec$ of $(\alpha,\delta)=$(00:21:30.3 $-$71:56:03), corresponding to the location of Bologna A \citep{2005A&A...435..871B}, are not included in our analysis. This does not affect our analysis as the background population is way outside the cluster center and its spatial extent is only 60$\arcsec$.  We show the respective fields of view in Figure \ref{Fig:FieldOfView}. CMDs for the space-based and ground-based are respectively shown in Figures \ref{Fig:CMDHubble} and \ref{Fig:CMDGroundBased}.

\section{Stellar Evolution Models}
We use the Yale Rotating Evolution Code \citep{2010arXiv1005.0423D} to compare our output parameters to theory for the RG and RGBB populations. Theoretical considerations for the HB population are taken from the literature \citep{1994A&A...285L...5R,1998A&A...334..953V,2001MNRAS.323..109G,2009A&A...499..835V,2010MNRAS.408..999D}.

\begin{figure}[H]
\includegraphics[totalheight=0.69\textheight]{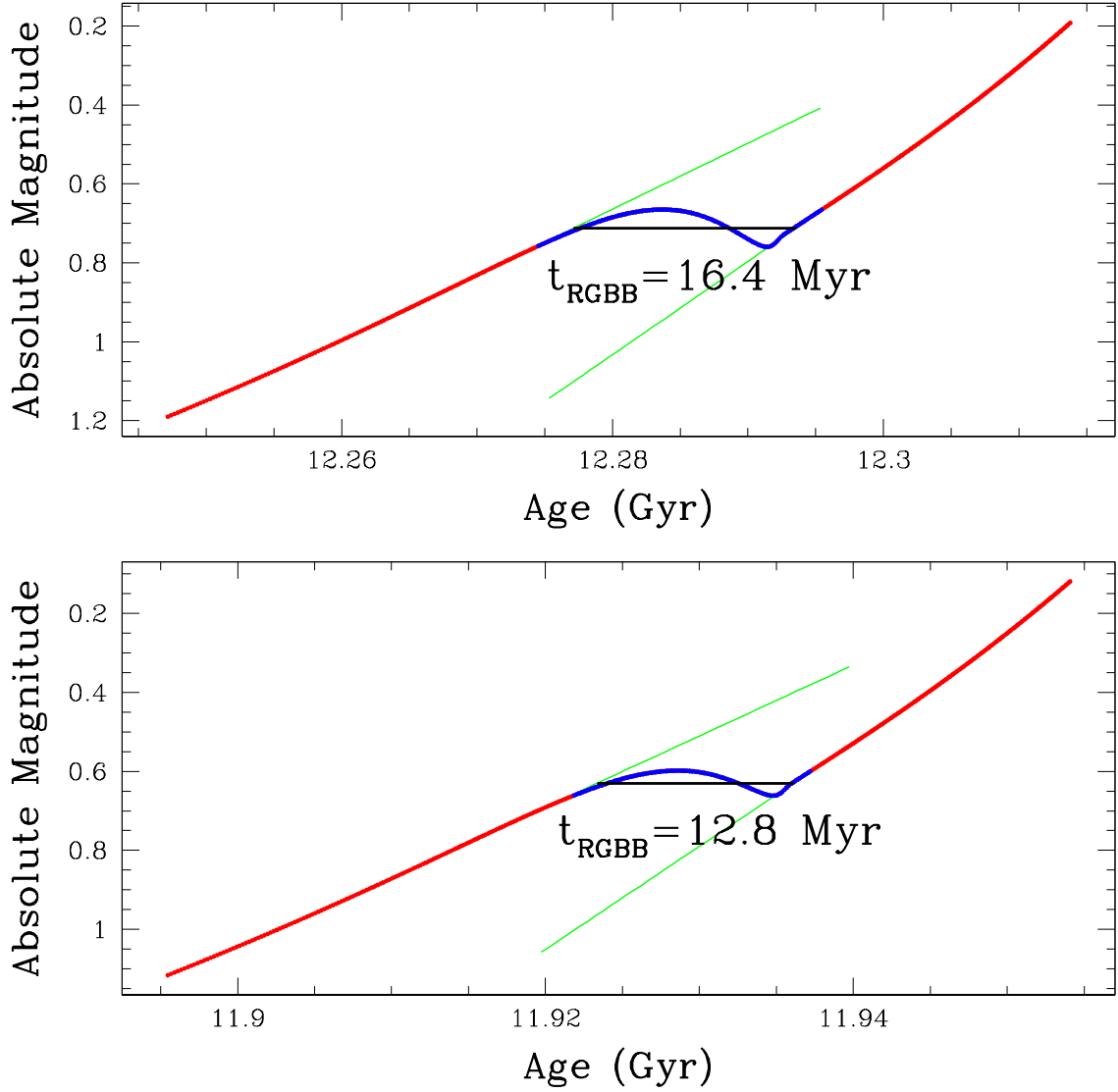}
\caption{Two representative stellar tracks for a 12 Gyr population with a metallicity [M/H]$ \approx -0.50$. Model points with a luminosity corresponding to the RGBB phase are plotted in blue, and the underlying red giant branch in red. TOP: Primordial helium content Y$=0.25$, the EW of the RGBB is 0.33 mag. BOTTOM: Primordial helium content Y$=0.28$, the EW of the RGBB is 0.28 mag. The EW is computed by multiplying the lifetime of the RGBB phase, shown by the length of the black horizontal line, by the average of the slopes of the luminosity evolution function just outside the RGBB, which are shown in green.}
\label{Fig:StellarTracks}
\end{figure}

At the expected metallicity ([M/H]$\sim -0.50$) and age ($\sim$12 Gyr) of 47 Tuc \citep{2008ApJ...684..326M,2010A&A...516A..55C}, we find that every 1\% increase in the initial helium abundance by total stellar mass yields a $\sim$10\% decrease in the lifetime of the RGBB, corresponding to a decrease of $\sim$0.02 mag in the EW. Two representative models are shown in Figure \ref{Fig:StellarTracks}. Within the models, we compute the EW of the RGBB by multiplying the lifetime of the RGBB by the average of the two slopes of magnitude versus time before and after the RGBB. The predicted stellar properties of the RGBB as a function of initial composition are summarized in Table 1. We note the helium-rich track has a higher initial [M/H] only because it has a lower initial hydrogen abundance -- the initial metallicity content by mass are the same. We also introduce the notation ${\delta}V_{RGBB}$ to refer to the difference in magnitudes between the brightest and faintest parts of the RGBB phase as predicted by stellar models.


\begin{table}[H]
\begin{center}
\caption{Predicted Stellar Properties of the RGBB}
    \begin{tabular}{ | l |  | l | l | }
     \hline
    -- & Model A & Model B  \\ \hline \hline \hline
    Y & 0.25 & 0.28  \\ \hline
    [M/H] & $-$0.52 & $-$0.50  \\ \hline
    Initial Mass & 0.90 M$_{\odot}$  &  0.86 M$_{\odot}$\\ \hline
    Age at RGBB & 12.28 Gyr & 11.93 Gyr \\ \hline
    Median RGBB brightness & 0.71(M$_{\rm{Bol}}$)  & 0.63 (M$_{\rm{Bol}}$) \\ \hline
    T$_{eff}$ at RGBB median brightness & 4536 K & 4534 K \\ \hline
    RGBB EW & 0.33 mag & 0.28 mag \\ \hline
    RGBB lifetime  & 16.4 Myr & 12.8 Myr \\ \hline
    Luminosity Evolution before RGBB & 16.71 mag Gyr$^{-1}$ & 17.99 mag Gyr$^{-1}$ \\ \hline
    Luminosity Evolution after RGBB & 23.65 mag Gyr$^{-1}$  & 26.11 mag Gyr$^{-1}$\\ \hline
    ${\delta}V_{RGBB} = V^{max}_{RGBB} - V^{min}_{RGBB}$  & 0.09 mag & 0.06 mag \\ \hline
    \end{tabular}
\end{center}
\label{Table:RGBBproperties}
\end{table}

\section{Red Giant Branch Bump Gradients in Brightness and Number Counts}
We show that the RGBB grows both fainter and more numerous relative to the RG stars with increased separation from the cluster center, and that the trends are both statistically significant. None of these trends are detected with significance in the inner $\sim$100$\arcsec$ of the cluster as traced by the space-based dataset, consistent with previous work \citep{1997AJ....114.1051B} and our measurements of the HB (discussed in a Section 5) that the two stellar populations are smoothly mixed in and near the cluster center. 

\subsection{Fitting for the RGBB}
In both datasets, we cut out a parallelogram around the red giant branch that keeps stars no more than 1.0 mag brighter nor 1.6 mag fainter than the RGBB in $V$. The range in magnitude is chosen to be as wide as possible while still excluding the HB and AGB stars. We show our color-magnitude cuts for the space-based and ground-based data in Figures \ref{Fig:CMDHubble} and \ref{Fig:CMDGroundBased}, respectively.

We fit for the RGBB and the RG using a combination of an exponential for the magnitude distribution of RG stars (equivalent to a power-law distribution in luminosity) and a Gaussian for the RGBB:
\begin{eqnarray}
N(m) = A\exp\biggl[B(V-V_{RGBB})\biggl] +\frac{N}{\sqrt{2\pi}\sigma}\exp \biggl[{-\frac{(V-V_{RGBB})^2}{2\sigma^2}}\biggl] \nonumber \\ \nonumber \\
N(m) = A\biggl\{\exp\biggl[B(V-V_{RGBB})\biggl] +\frac{EW}{\sqrt{2\pi}\sigma}\exp \biggl[{-\frac{(V-V_{RGBB})^2}{2\sigma^2}}\biggl]\biggl\}
\label{EQ:Parameters},
\end{eqnarray}
where $V_{RGBB}$ is the peak magnitude of the RGBB, $\sigma$ is the dispersion, $N$ is the number of RGBB stars, and $A$ and $B$ are the normalization and scale of the exponential. As in \citet{2010arXiv1011.4293N}, we define the equivalent width of the RGBB, EW, to be the ratio of the number of RGBB stars to the number density of RG stars at the magnitude of the RGBB. In this parametrization $EW = N/A$. We use Markov Chain Monte Carlo (MCMC) to obtain the maximum likelihood values for the parameters. For each value of the parameters tested by the MCMC, we compute the log-likelihood $\ell$:
\begin{equation}
\ell = \sum_{i}^{N_{obs}}ln\biggl[N(m/A,B,EW,\sigma,V_{RGBB})\biggl],
\end{equation}
where $N_{obs}$ is the total number of stars, and the parameter $A$ is selected in each run of the MCMC such that the integral of the function $N(m)$ over the magnitude range is equal to $N_{obs}$. We do not fit an RGBB model to the data by first binning it, but it can be shown that this method is equivalent to binning data in the limit of infinitesimal bin widths.

\subsection{Decreased Brightness and Increased Equivalent Width for RGBB Stars Farther From the Cluster Core}
Fitting for gradients in the brightness of the RGBB $V_{RGBB}$ and the number counts parameter EW with separation from the cluster center yields statistically significant improvements in the fit for the ground-based dataset, with no significant improvement in the space-based dataset toward the cluster core. 

We fit for a gradient using the following extension to our parametrization:
\begin{eqnarray}
V_{RGBB} = V_{RGBB,0} + \frac{dV_{RGBB}}{d\log(r)}\times \biggl[\log(r) - \overline{ \log(r)}\biggl]  \nonumber \\ \nonumber \\
EW = EW_{0} + \frac{dEW}{d\log(r)}\times \biggl[\log(r) - \overline{ \log(r)}\biggl]
\label{EQ:Parameters2},
\end{eqnarray}
where $r = (R_{i}-R_{CC})$ is the separation in arcseconds of the location of the i\textit{th} star, $R_{i}$ from the cluster center $R_{CC}$, taken here as being $(\alpha, \delta) =$ (00:24:05.4, $-$72:04:53), and is taken from Vizier \citep{2000A&AS..143...23O}. The mean of the logarithmic separation for all the RG+RGBB stars used to construct the fit is designated $\overline{ \log(r)}$. We tested parametrizations of both separation, squared separation and the log of separation and found that using the logarithmic separation from the cluster center yielded the largest improvement in the fit as measured by ${\Delta}\ell$. 

\begin{figure}[H]
\includegraphics[totalheight=0.78\textheight]{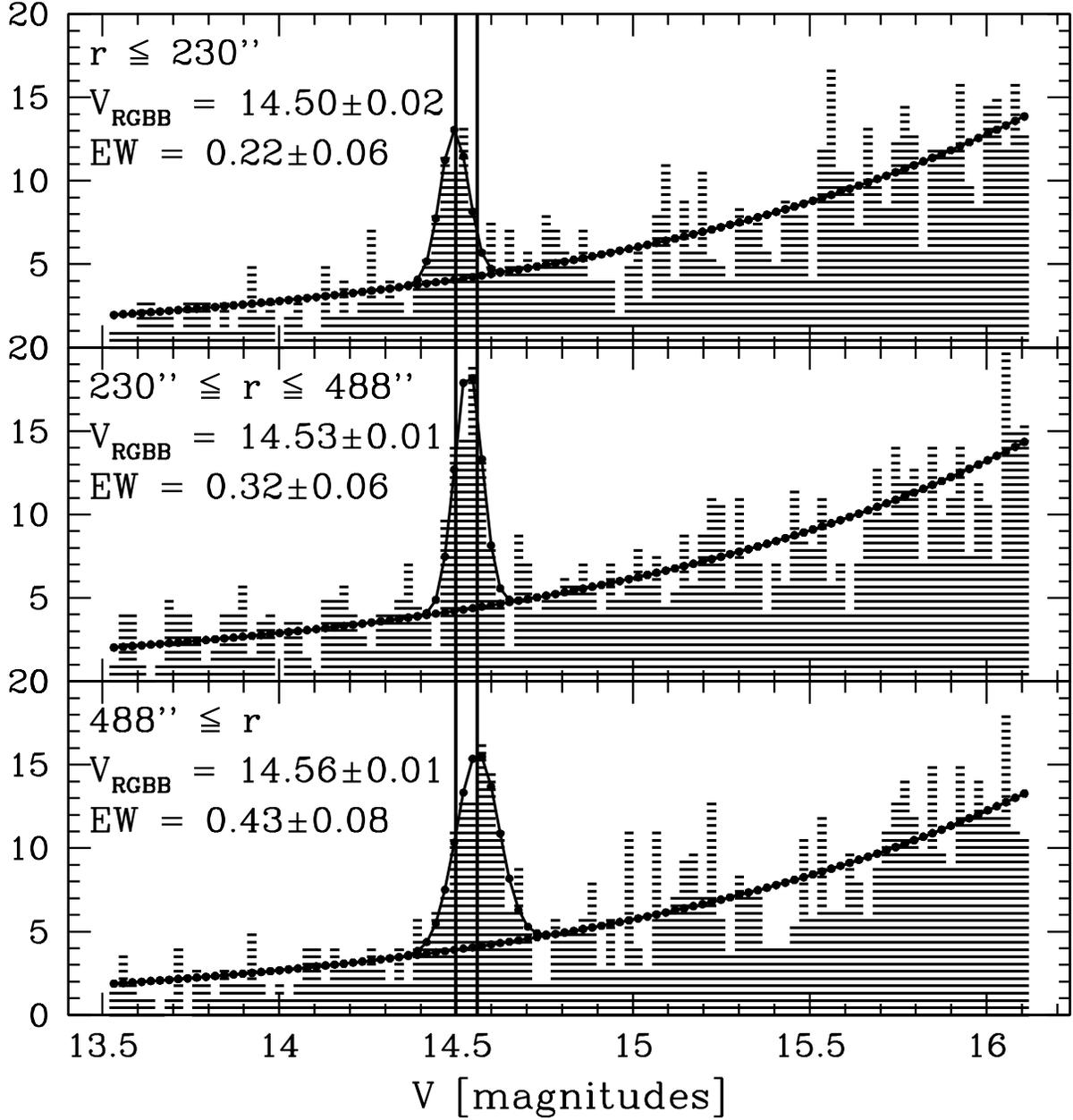}
\caption{The RG+RGBB stellar population of the groundbased dataset is split up into three equal subpopulations, sorted by radial separation from the center of the cluster. The trends expected in the presence of a helium gradient, that of increasing EW and declining brightness, are both observed. We have fixed the exponential slope of the power law to $B=0.76$, as is obtained when fitting to the entire cluster. This does not affect the parameter $V_{RGBB}$ but it does slightly reduce the trend in EW. Vertical lines are drawn in each panel corresponding to the peaks of the inner and outer RGBB peaks, at $V=14.50$ and $V=14.56$.}
\label{Fig:RGBBgradients}
\end{figure}

Allowing gradients for both $V_{RGBB}$ and EW yields ${dV_{RGBB}}/{d\log(r)} = (0.083 \pm 0.023)$ mag dex$^{-1}$, and ${dEW}/{d\log(r)} = (0.27 \pm 0.12)$ mag dex$^{-1}$. These two gradients each retain nearly identical values when the other is fixed to being zero and can therefore be considered independent. Both gradients go in the direction expected from stellar theory in the presence of a helium gradient. The model predictions shown in Figure \ref{Fig:StellarTracks} predict that for a 12 Gyr population with a metallicity [M/H]$ \approx -0.50$, the brightness should increase by $\sim$0.05 and the EW should decrease by $\sim$0.07 mag and as the initial helium abundance is increased from Y=0.25 to Y=0.28.  Since the total span of the ground-based dataset in arcseconds is approximately 1 dex, the measured gradients are a little higher than the theoretical expectations. A graphical representation of these gradients is shown in Figure \ref{Fig:RGBBgradients}

\section{Horizontal Branch Gradients in Color and Brightness}
The trends in color and brightness with respect to the distance to the cluster center for HB stars are compatible with an even mixing of the two stellar generations in the cluster center and with the helium-enhanced second generation decreasing more rapidly at larger radii. We discuss four possible contaminating effects and find that none of them have the same predicted trends as are observed. 

As there is no definitive way to cleanly and completely separate horizontal branch stars from background contamination stars and AGB stars, we utilize visually satisfactory color-magnitude boxes in the space-based and ground-based datasets, respectively shown in Figures  \ref{Fig:CMDHubble} and \ref{Fig:CMDGroundBased}. In the space-based data, we tabulate 545 stars with $0.82<(V-I)<0.98$ and $13.8<V<14.2$, whereas in the ground-based dataset we tabulate 771  stars with $0.7<(B-V)<0.9$ and $13.8<V<14.2$. For both datasets, we compute the correlation between brightness and the log of the separation from the cluster center.

In the ground-based data, the HB stars become fainter and redder with increased separation from the cluster center. The brightness gradient is $(0.072 \pm 0.009)$ mag dex$^{-1}$ in $V$, and the color gradient is  $(0.021 \pm 0.005)$ mag dex$^{-1}$ in $(B-V)$. Using only those sources that have at least 5 measurements in each filter, we also obtain color gradients of $(0.064 \pm 0.009)$ mag dex$^{-1}$ in $(B-I)$ and $(0.078 \pm 0.013)$ mag dex$^{-1}$ in $(U-V)$. The gradients go in the same direction as expected if a helium-enriched, second generation of GC cluster exists in 47 Tuc with a more central concentration. The HB stars in the space-based dataset, toward the cluster center, has  a slight trend of getting redder with increased separation from the cluster center, but it is a very weak trend, 1.3$\sigma$, and is without a corresponding trend in brightness. 

\begin{figure}[H]
\begin{center}
\includegraphics[totalheight=0.6\textheight]{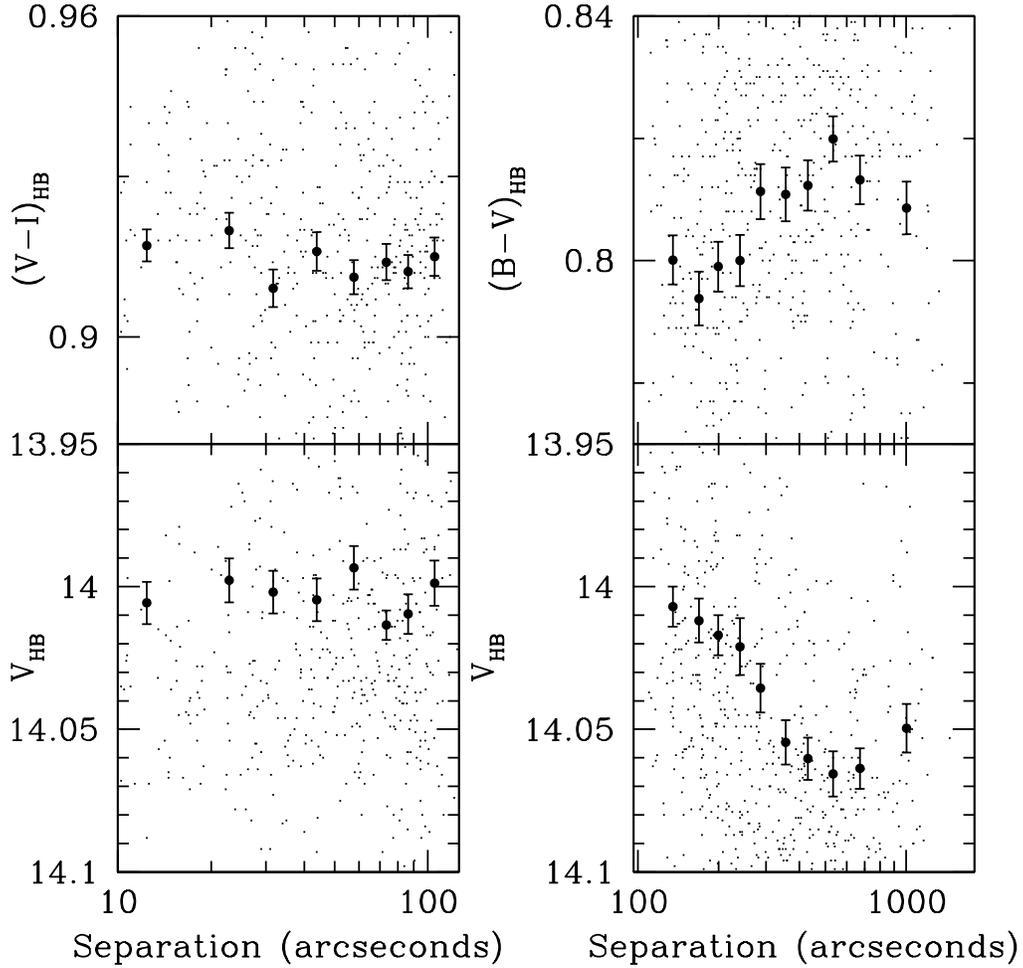}
\end{center}
\caption{Color and brightness of the HB stars binned as a function of separation from the cluster center. Left-hand panels show the data for the cluster-core and right-hand panels for the rest of the cluster. Top panels show the color of the HB star, and bottom panels show the brightness in $V$. The thick black points are the means and the error bars are the standard error in the mean for subsamples of $\sim$70 HB stars binned in separation from the cluster center. The more numerous, less sharp points correspond to individual HB stars.}
\label{Fig:RHBbrightness}
\end{figure}

We show the colors and brightness for HB stars as a function of separation from the cluster center in Figure \ref{Fig:RHBbrightness}. The profile is one of even mixing between the two populations within the inner $\sim$200$\arcsec$, with the brighter HB stars falling off in relative numbers between $\sim$200 and $\sim$400$\arcsec$.  That the color and brightness profiles show the same structure with respect to separation from the cluster center is evidence that the trends are genuine attributes of the cluster stars rather than statistical fluctuations. 

The potential contamination effects in this comparison would not go in the same direction as the measured gradient. First, both signals go in the \textit{opposite} direction to the weak signal expected from the known CNO variation in the cluster. As stars in the cluster center have higher CNO, they ought to be slightly ($\sim$0.01 mag) redder and fainter with decreased distance to the cluster center. We refer the reader to Figure 1 of \citet{2010MNRAS.408..999D}. Second, this result cannot be explained by having the second population be substantially younger. A younger HB population at this metallicity would indeed be brighter, but it would also be redder rather than bluer \citep{2001MNRAS.323..109G}. A third possible effect is that of binary interactions. Stars closer to the cluster center will be more likely to have a binary companion \citep{2007MNRAS.380..781S,2007ApJ...665..707H}, which implies increased mass loss during the red giant branch. That would make HB stars bluer with decreasing distance to the cluster center as we observe, but it would also make them fainter.  In their recent dynamical investigation of the cluster, \citet{2010MNRAS.tmp.1747G} found that the fraction of stars in binary pairs should level off after a $\sim$100$\arcsec$ separation from the cluster center. Hence, if binary interactions were the cause of this effect they would induce a much tighter radial profile. Lastly, the fact that each of the color gradients is $\sim$3$\times$ larger than that detected for the RG stars at the level of the RGBB (discussed in section 6) contradicts the hypothesis that the HB gradients would be due to a reddening or blending gradient. 

\section{Note Regarding a Small Color Gradient on the Red Giant Branch}
We find evidence for a small color gradient in the cluster RG stars that is not a predicted outcome of the models in the presence of a helium gradient. We compute the least-squares relation between the measured colors of all the RG+RGBB stars and their separation from the center, $(\log(r) - \overline{ \log(r)})$, using those stars that have at least 5 measurements in each filter and that have a brightness within 0.5 mag of $V_{RGBB}$.  The measured slope in $(B-V)$ color is a miniscule $(0.005 \pm 0.002)$ mag dex$^{-1}$ -- effectively zero since the total extent of the ground-based dataset is $\sim$1.1 dex. However, the gradient measured in $(B-I)$ color is  $(0.027 \pm 0.003)$ mag dex$^{-1}$, and is $(0.028 \pm 0.007)$ mag dex$^{-1}$ in $(U-V)$ color. Each of these gradients is significantly smaller than that for the HB stars, discussed in Section 5. 

Our model results, shown in Table \ref{Table:RGBBproperties}, predict that there should be no measurable color gradient due to temperature in the cluster at the level of the RGBB if we are only dealing with two equal-age, equal-metallicity populations differing only in their initial helium abundance by a small amount (${\Delta}Y \sim$ 0.03). The presence of these small gradients indiciates there may be an additional factor at play, or that helium-enriched red giants might be made slightly bluer than predicted by models. The color variation would be consistent with the stars nearer the center either having a \textit{lower} metallicity, ${\delta}$[Fe/H] $\approx$ 0.05 dex, or a temperature colder by ${\delta}$T $\approx$ 17 Kelvin \citep{1999A&AS..140..261A}.

\section{Discussion}

We have found $\sim$3.6$\sigma$ and $\sim$2.3$\sigma$ detections that the cluster RGBB stars become fainter and more numerous with increasing distance from the cluster center, and that the HB becomes fainter and redder at the $\sim$7.7$\sigma$ and $\sim$4.1$\sigma$ levels respectively. These four independent effects are easily explained by stellar theory if there is a second generation of stars in 47 Tuc that is helium-enhanced and more centrally concentrated.  \citet{1997AJ....114.1051B}, in his analysis of CN-band indices in 283 cluster giants, found a similar radial profile for 47 Tuc. The ratio of CN-strong to CN-weak cluster stars was approximately equal ($\sim$ 1.8) interior to 10$\arcmin$, and then dropped steeply for sources separated from the cluster center by 10-20$\arcmin$. If this is due to dynamical segregation between the two generations, then the thickness of the main-sequence observed by \citet{2009ApJ...697L..58A} in the cluster core should not be observable in the cluster outskirts. This will prove a difficult measurement to make since the surface density of stars on the sky drops steeply at these distances. Additional ground-based data could also prove useful. We estimate that there are $\sim$30 cluster RGBB stars outside the range of our observations, these should be slightly fainter and with a higher equivalent width than those measured thus far. The remaining $\sim$100 HB stars should also be fainter and redder than those within 200$\arcsec$ of the cluster center. 

In spite of the evidence for an overall gradient, it is interesting that we do not find evidence of a gradient within the cluster center as traced by the \textit{HST} dataset. We discuss four possible explanations. One way to explain this effect is to have the two stellar populations evenly mixed within some radius, this would lead to a flattening of all of the indicators we tested, as observed. Unfortunately, only $\sim$20\% of the RGBB and HB stars in the space-based dataset are contained within the King radius of the cluster, estimated to be 20.84$\arcsec$ \citep{2006ApJS..166..249M}. No significant trend is detected with the stars outside that radius and within the space-based dataset, indicating they may initially fall off in number density at similar rates for $r \gtrapprox r_{king}$. A second way would be if 47 Tuc indeed has a third stellar generation, as suggested by \citet{2010MNRAS.408..999D}. These stars, comprising $\sim$10\% of the cluster population, would be formed from helium-enriched AGB ejecta that was 50\% diluted by pristine material, and would as such have intermediate initial helium abundance relative to the first and second generation. If they are the most centrally concentrated stellar generation, they would cause population tracers of helium abundance to level off. A third possibility is the role of binary interactions. In their dynamical model of the cluster, \citet{2010MNRAS.tmp.1747G} predicted that the binary fraction should be steeply varying for stars with projected separations of 10-100$\arcsec$ from the cluster center, levelling off for larger separations. That range of separations is precisely that traced by the \textit{HST} dataset, and if the dynamical predictions are correct the role of binary evolution may be ``blurring'' the signals in our color-magnitude tracers for the HB. However, since the RGBB takes place very early in the ascent of a red giant branch, we do not expect the RGBB to be affected by binary evolution, thereby limiting the explanatory power of this third hypothetical effect.

Systematic errors in either or both of the two datasets are a concern, and it would be preferable to have a deep, high-precision uniform-photometry dataset over the entire cluster, but such ideal data are currently unavailable so we have combined the best photometry available. Within the ground-based dataset, it is very difficult to conceive of the broad range of systematics necessary to produce our signals. The decreasing brightness trend for the HB, $(0.072 \pm 0.009)$ mag dex$^{-1}$, is comparable to that of the RGBB, $(0.083 \pm 0.023)$ mag dex$^{-1}$. The similarity may be suggestive of a systematic,  however, we also detect a significant color trend with the HB of $(0.021 \pm 0.005)$ mag dex$^{-1}$ with no analog for the RGBB. The RGBB shows no color gradient in the ground-based dataset regardless of whether or not we allow for gradients in $V_{RGBB}$ and EW. As a further constraint on any potential systematic, the parameter we used to model the RGBB, EW, is independent of any continuous photometric completeness function since the RGBB stars will necessarily have the same probability of detection as the other RG stars of the same brightness, which makes this parametrization very robust. 

This is, as far as we know, the first empirical support for the stellar theory prediction that the RGBB lifetime should shorten as initial helium abundance is enhanced  \citep{2001ApJ...546L.109B,2010ApJ...712..527D,2010arXiv1011.4293N}. This opens the prospect that the RGBB can now provide a diagnostic to measure the helium abundance in populations such as dwarf galaxies and galactic bulges. This diagnostic should be more effective in metal-rich populations ([M/H] $\gtrsim -0.5$) because there are more RGBB stars at higher metallicity (for fixed age and helium enrichment) rendering any given fluctuation more statistically significant. However, recent research has demonstrated that theory may inaccurately overestimate the RGBB luminosity by $\sim$0.2 mag  \citep{2011A&A...527A..59C}, which was estimated by comparing the difference in brightness between the RGBB and the main-sequence turn-off for a sample of 15 Galactic globular clusters, and that predicted given the estimated ages and measured metallicities of the cluster. If there are small systematic errors in the theoretical predictions for RGBB luminosity, there may also be errors in the predictions for the RGBB lifetime. On that note, our brightness gradients for the HB, $(0.072 \pm 0.009)$ mag dex$^{-1}$, and for the RGBB, $(0.083 \pm 0.023)$ mag dex$^{-1}$, yield a ratio of $dV_{RGBB}/dV_{HB}$ = $(1.15 \pm 0.35)$. This is unfortunately insufficiently precise to test a stellar theory prediction of \citet{1997MNRAS.285..593C}. Their models predicted that that the RGBB luminosity should respond more steeply to varying initial helium abundance than the ZAHB luminosity, and as such the difference in their V-band brightness should decrease by $\sim$0.011 mag for every increase of 0.01 in Y, suggesting a ratio $dV_{RGBB}/dV_{HB} \sim 2$. A more precise comparison is within reach if uniform photometry is obtained over the entirety of the cluster.

\acknowledgments
DMN and AG were partially supported by the NSF grant AST-0757888.

\end{document}